\documentclass[fleqn,12pt]{article}
\usepackage{espcrc1m}
\usepackage{graphicx}
\usepackage{latexsym}

\newcommand{\beq}{\begin{equation}}
\newcommand{\eeq}{\end{equation}}
\newcommand{\bqa}{\begin{eqnarray}}
\newcommand{\eqa}{\end{eqnarray}}

\title{Thermalization and plasma instabilities}

\author{Michael Strickland
\address[FRANKFURT]{Frankfurt Institute for Advanced Studies \\
  Johann Wolfgang Goethe - Universit\"at Frankfurt \\
  Max-von-Laue-Stra\ss{}e~1 \\
  D-60438 Frankfurt am Main \\ 
  Germany \\}
}

\begin{document}

\maketitle

\abstract{
I review recent analytical and numerical advances in the study of non-equilibrium 
quark-gluon plasma physics.  I concentrate on studies of the 
dynamics of plasmas which are locally anisotropic in momentum space.  In 
contrast to locally isotropic plasmas such anisotropic plasmas have a spectrum 
of soft unstable modes which are characterized by exponential growth of 
transverse (chromo)-magnetic fields at short times.  Parametrically the 
instabilities provide the fastest method for generation of soft background 
fields and dominate the short-time dynamics of the system.
}

\section{Introduction}

One of the most important open questions emerging from the
Relativistic Heavy-Ion Collider (RHIC) program is how quickly, and by
what process, the stress-energy tensor of the high-density QCD matter
produced in the central rapidity region becomes isotropic. This is
important, for example, to determine the maximal temperature achieved
in such collisions, which will also be pursued in the near future at
even higher energies at the CERN Large Hadron Collider (LHC).
One of the chief obstacles to thermalization in ultrarelativistic
heavy-ion collisions is the rapid longitudinal expansion of the
central rapidity region. If the matter expands too quickly then there
will not be sufficient time for its constituents to interact and
thermalize.  In the absence of interactions, the longitudinal
expansion causes the system to become much colder in the longitudinal
direction than in the transverse directions, corresponding to $\langle
p_L^2 \rangle \ll \langle p_T^2\rangle$ in the local rest frame.  One
can then ask how long it would take for interactions to restore
isotropy in momentum space.

High-energy heavy-ion collisions release a large amount of partons
from the wavefunctions of the colliding nuclei. Partons with very
large transverse momenta originate from high-$Q^2$ hard interactions,
while partons with transverse momenta below the so-called saturation
momentum $Q_s$ (given by the square root of the color charge density
per unit area in the incoming nuclei) are much more abundant and are
better viewed as a classical non-Abelian
field~\cite{Mueller:2002kw,Iancu:2003xm,McLerran:2005kk}.  At high
energies, $Q_s\gg\Lambda_{\rm QCD}$ sets a semi-hard scale which
allows for a weak-coupling treatment of the early-stage
dynamics. However, due to the large occupation number of the soft
modes below $Q_s$ (the classical color field has strength
$F_{\mu\nu}\sim1/g$), the problem is nevertheless non-perturbative.

The evolution following the initial impact was among the questions
which the so-called ``bottom-up scenario''~\cite{Baier:2000sb}
attempted to answer. For the first time, it addressed the dynamics of
soft modes (``fields'') with momenta much below $Q_s$ coupled to the
hard modes (``particles'') with momenta on the order of $Q_s$ and
above. However, it has emerged recently that one of the assumptions
made in this model is not correct.  The debate centers around the very
first stage of the bottom-up scenario in which it was assumed that (a)
collisions between the high-momentum (or hard) modes were the driving
force behind isotropization and that (b) the low-momentum (or soft)
fields act only to screen the electric interaction. In
doing so, the bottom-up scenario implicitly assumed that the
underlying soft gauge modes behaved the same in an anisotropic plasma
as in an isotropic one.  However, it turns out that in anisotropic
plasmas, within the so-called ``hard loop approximation'' (see below),
the most important collective mode corresponds to an instability to
transverse magnetic field fluctuations
\cite{Mrowczynski:1993qm,Mrowczynski:1994xv,Mrowczynski:1996vh}.
Recent works have shown that the presence of these instabilities is
generic for distributions which possess a momentum-space anisotropy
\cite{Romatschke:2003ms,Arnold:2003rq,Romatschke:2004jh} and have
obtained the full hard-loop (HL) action in the presence of an
anisotropy \cite{Mrowczynski:2004kv}.  Another important development
has been the demonstration that such instabilities exist in numerical 
solutions to classical Yang-Mills fields in an expanding geometry 
\cite{Romatschke:2005ag,Romatschke:2005pm,Romatschke:2006nk}.

Recently there have been significant advances in the
understanding of non-Abelian soft-field dynamics in anisotropic
plasmas within the HL
framework~\cite{Arnold:2005vb,Rebhan:2005re,Romatschke:2006wg}.  The
HL framework is equivalent to the collisionless Vlasov theory of
eikonalized hard particles, i.e.\ the particle trajectories are
assumed to be unaffected (up to small-angle scatterings with $\theta \sim g$)
by the induced background field. It is
strictly applicable only when there is a large scale separation
between the soft and hard momentum scales.  Even with these simplifying assumptions,
HL dynamics for self-interacting gauge fields is non-trivial due to
the presence of non-linear interactions which can act to regulate
unstable growth.  These non-linear interactions become important when
the vector potential amplitude is on the order of $A_{\rm non-Abelian}
\sim p_{\rm s}/g \sim (g p_h)/g$, where $p_h$ is the characteristic
momentum of the hard particles and $p_s$ is the characteristic soft
momentum of the fields.  In QED there is no
such complication and the fields grow exponentially until $A_{\rm
Abelian} \sim p_h/g$ at which point the hard particles undergo
large-angle deflections by the soft background field invalidating the
assumptions underpinning the hard-loop effective action.

Recent numerical studies of HL gauge dynamics for SU(2) gauge theory
indicate that for {\em moderate} anisotropies the gauge field dynamics
changes from exponential field growth indicative of a conventional
Abelian plasma instability to linear growth when the vector potential
amplitude reaches the non-Abelian scale, $A_{\rm non-Abelian} \sim
p_{\rm h}$ \cite{Arnold:2005vb,Rebhan:2005re}. This linear growth
regime is characterized by a cascade of the energy pumped into the
soft modes by the instability to higher-momentum plasmon-like modes
\cite{Arnold:2005ef,Arnold:2005qs}. These results indicate that there
is a fundamental difference between Abelian and non-Abelian plasma
instabilities.  However, even with this new understanding the HL
framework relies on the existence of a large separation between the
hard and soft momentum scales by design.  One would like to know what
happens when the scale separation between hard and soft modes is not
very large or when the initial fields have large (non-linear)
amplitudes which is seemingly the situation faced in real
experiments. In this case one is naturally led to consider instead numerically
solving the Wong-Yang-Mills (WYM) equations \cite{Dumitru:2005gp,Dumitru:2006pz}.

\section{Anisotropic Gluon Polarization Tensor}

In this section I consider a quark-gluon plasma with a parton distribution function
which is decomposed as
\bqa
f({\bf p}) \equiv 2 N_f \left(n({\bf p}) + \bar n ({\bf p})\right) + 4 N_c n_g({\bf p}) \; ,
\label{distfncs}
\eqa
where $n$, $\bar n$, and $n_g$ are the distribution functions of quarks, anti-quarks, 
and gluons, respectively, and the numerical coefficients collect all appropriate
symmetry factors.
Using the result of Ref.~\cite{Romatschke:2003ms} the spacelike 
components of the high-temperature gluon self-energy for gluons with soft 
momentum ($k \sim g T$) can be written as
\begin{equation}
\Pi^{i j}(K) = - \frac{g^2}{2} \int \frac{d^3{\bf p}}{(2\pi)^3} v^{i} \partial^{l} f({\bf p})
\left( \delta^{j l}+\frac{v^{j} k^{l}}{K\cdot V + i \epsilon}\right) \; ,
\label{selfenergy2}
\end{equation}
%
where the parton distribution function $f({\bf p})$ is, in principle, 
completely arbitrary.  In what follows we will assume that $f({\bf p})$ can be 
obtained from an isotropic distribution function by the rescaling of only one 
direction in momentum space.  

In practice this means that, given any isotropic 
parton distribution function $f_{\rm iso}(p)$, we can construct an 
anisotropic version by changing the argument of the isotropic distribution 
function, 
$f({\bf p}) = \sqrt{1+\xi} \, f_{\rm iso}\left(\sqrt{{\bf p}^2+\xi({\bf p}\cdot{\bf \hat n})^2}\right)$,
where the factor of $\sqrt{1+\xi}$ is a normalization constant which ensures 
that the same parton density is achieved regardless of the anisotropy introduced, 
${\bf \hat n}$ is the direction of the anisotropy, and $\xi>-1$ is an adjustable 
anisotropy parameter with $\xi=0$ corresponding to the isotropic case. 
Here we will concentrate on $\xi>0$ which corresponds to a 
contraction of the distribution along the ${\bf \hat n}$ direction since this is 
the configuration relevant for heavy-ion collisions at early times, namely
two hot transverse directions and one cold longitudinal direction.

Making a change of variables 
in (\ref{selfenergy2}) it is possible to integrate out the $|p|$-dependence 
giving~\cite{Romatschke:2003ms}
\begin{equation}
\Pi^{i j}(\omega/k,\theta_n) = \mu^2 \int \frac{d \Omega}{4 \pi} v^{i}%
\frac{v^{l}+\xi({\bf v}\cdot\hat{\bf n}) \hat{n}^{l}}{%
(1+\xi({\bf v}\cdot\hat{\bf n})^2)^2}
\left( \delta^{j l}+\frac{v^{j} k^{l}}{K\cdot V + i \epsilon}\right) ,
\end{equation}
%
where $\cos\theta_n \equiv \hat{\bf k}\cdot\hat{\bf n}$ and $\mu^2 \equiv 
\sqrt{1+\xi}\;m_D^2>0$. The isotropic Debye mass, $m_D$, depends on 
$f_{\rm iso}$.  In the case of pure-gauge QCD with an equilibrium 
$f_{\rm iso}$ we have $m_D = g T$.

The next task is to construct a tensor basis for the 
spacelike components of the gluon self-energy and propagator.  We therefore need 
a basis for symmetric 3-tensors which depend on a fixed anisotropy 3-vector 
$\hat{n}^{i}$ with $\hat{n}^2=1$.  This can be achieved with the following four component 
tensor basis: $A^{ij} = \delta^{ij}-k^{i}k^{j}/k^2$, $B^{ij} = k^{i}k^{j}/k^2$, 
$C^{ij} = \tilde{n}^{i} \tilde{n}^{j} / \tilde{n}^2$, and $D^{ij} = 
k^{i}\tilde{n}^{j}+k^{j}\tilde{n}^{i}$ with $\tilde{n}^{i}\equiv A^{ij} \hat{n}^{j}$. 
Using this basis we can decompose the self-energy into four structure functions 
$\alpha$, $\beta$, $\gamma$, and $\delta$ as ${\bf \Pi}= \alpha\,{\bf A} + 
\beta\,{\bf B} + \gamma\,{\bf C} + \delta\,{\bf D}$.  Analytic 
integral expressions for $\alpha$, $\beta$, $\gamma$, and $\delta$ can be found 
in Ref.~\cite{Romatschke:2003ms} and \cite{Romatschke:2004jh}.

\section{Collective Modes}

As shown in Ref.~\cite{Romatschke:2003ms} this tensor basis allows us to express the 
propagator in terms of the following three functions
\bqa
\Delta_\alpha^{-1}(K) &=& k^2 - \omega^2 + \alpha \; , \label{propfnc1} \nonumber \\
\Delta_\pm^{-1}(K) &=& \omega^2 - \Omega_\pm^2 \; , \nonumber
\label{propfnc2}
\eqa
where $ 2 \Omega_{\pm}^2 = \bar\Omega^2 \pm \sqrt{\bar\Omega^4- 4 
((\alpha+\gamma+k^2)\beta-k^2\tilde n^2\delta^2) }$
and $\bar\Omega^2 = \alpha+\beta+\gamma+k^2$.  

Taking the static limit of these 
three propagators we find that there are three mass scales:  $m_\pm$ and $m_\alpha$. In 
the isotropic limit, $\xi\rightarrow0$, $m_\alpha^2=m_-^2=0$ and $m_+^2 = m_D^2$.  
However, for $\xi>0$ we find that $m_\alpha^2<0$ for 
all $\mid\!\!\theta_n\!\!\mid\,\neq\pi/2$ and $m_-^2<0$ for 
all $\mid\!\!\theta_n\!\!\mid\,\leq\pi/4$.  Note also that for $\xi>0$ both $m_\alpha^2$ and $m_-^2$
have there largest negative values at $\theta_n=0$ where they are equal.

The fact that for $\xi>0$ both $m_\alpha^2$ and $m_-^2$ can be negative 
indicates that the system is unstable to both magnetic and electric fluctuations 
with the fastest growing modes focused along the beamline ($\theta_n=0$).   In 
fact it can be shown that there are two purely imaginary solutions to each of the 
dispersions relations $\Delta_\alpha^{-1}(K)=0$ and $\Delta_- ^{-1}(K)=0$ with the 
solutions in the upper half plane corresponding to unstable modes.  We can 
determine the growth rate for these unstable modes by taking $w\rightarrow 
i\Gamma$ and then solving the resulting dispersion relations for $\Gamma(k)$.

\begin{figure}
$\;\;\;\;
$\begin{minipage}[t]{6.5cm}
\vspace{-5.8cm}
 In Figure~\ref{fig1} I plot the instability growth rates, $\Gamma_\alpha$ and $\Gamma_-$, as a function
 of wave number for $\xi=10$ and $\theta_n=\pi/8$.  Note that both growth rates vanish at $k=0$ and
 have a maximum $\Gamma_*\sim\mu/10$ at $k_*\sim\mu/3$.  The fact that they have a maximum
 means that at early times the system will be dominated by unstable modes with spatial frequency
 $1/k_*$. 
\end{minipage}
$\;\;\;\;\;\;\;\;\;$
\begin{minipage}[t]{8cm}
\includegraphics[width=7.6cm]{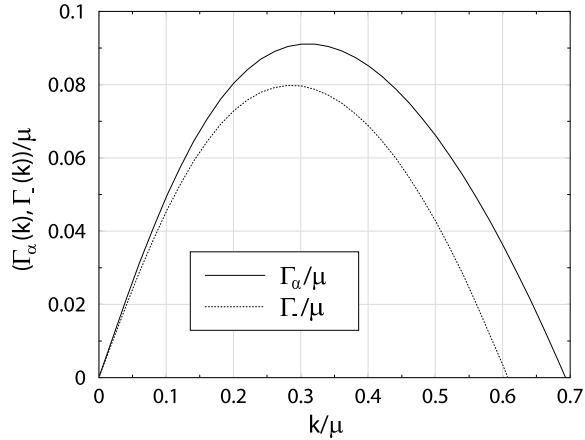}
\end{minipage}
\vspace{-8mm}
\caption{Instability growth rates as a function of wave number for $\xi=10$ and $\theta_n=\pi/8$.}
\label{fig1}
\end{figure}

\section{Discretized Hard-Loop Dynamics}%

It is possible to go beyond an analysis of gluon polarization tensor to a
full effective field theory for the soft modes and then solve this numerically.  
The effective field 
theory for the soft modes that is generated by integrating out the hard plasma 
modes at one-loop order and in the approximation that the amplitudes of the soft 
gauge fields obey $A_\mu \ll |\mathbf p|/g$ is that of gauge-covariant 
collisionless Boltzmann-Vlasov equations \cite{Blaizot:2001nr}. In equilibrium, the 
corresponding (nonlocal) effective action is the so-called hard-thermal-loop 
effective action which has a simple generalization to plasmas with anisotropic 
momentum distributions \cite{Mrowczynski:2004kv}. For the general non-equilibrium
situation the resulting equations of motion are
\begin{eqnarray}
D_\nu(A) F^{\nu\mu} &=& -g^2 \int {d^3p\over(2\pi)^3} {1\over2|\mathbf p|} \,p^\mu\, 
						 {\partial f(\mathbf p) \over \partial p^\beta} W^\beta(x;\mathbf v) \, , \nonumber \\
F_{\mu\nu}(A) v^\nu &=& \left[ v \cdot D(A) \right] W_\mu(x;\mathbf v) \, , 
\label{eom}
\end{eqnarray}
where $f$ is a weighted sum of the quark and gluon distribution 
functions \cite{Mrowczynski:2004kv} and $v^\mu\equiv p^\mu/|\mathbf p|=(1,\mathbf v)$.

These equations include all hard-loop resummed propagators and vertices and are 
implicitly gauge covariant.  At the expense of introducing a continuous set of auxiliary fields 
$W_\beta(x;\mathbf v)$ the effective field equations are also local.  These equations 
of motion are then discretized in space-time and ${\mathbf v}$, and solved 
numerically.  The discretization in ${\mathbf v}$-space corresponds to including 
only a finite set of the auxiliary fields $W_\beta(x;\mathbf v_i)$ with $1 \leq 
i \leq N_W$. For details on the precise discretizations used see 
Refs.~\cite{Arnold:2005vb,Rebhan:2005re}.

\begin{figure}[t]
\vspace{4mm}
\centerline{
\includegraphics[width=9.1cm]
{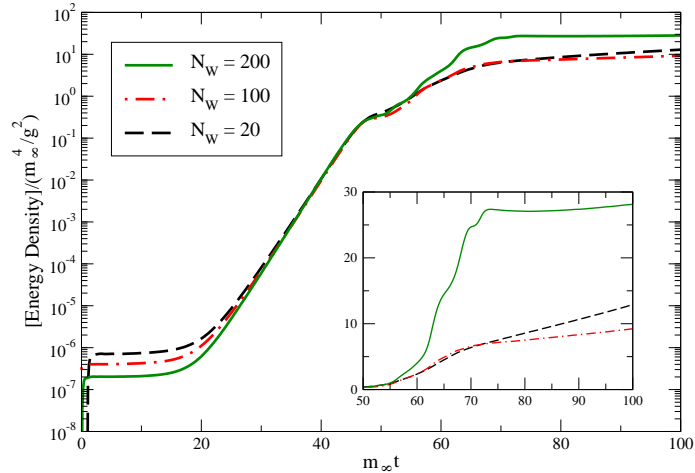}
}
\vspace{-7mm}
\caption{
Comparison of the energy transferred from hard to soft scales, $\mid\!\!\mathcal 
E({\rm HL})\!\!\mid$, for 3+1-dimensional simulations with $N_w=20,100,200$ 
on $96^3, 88^3, 69^3$ lattices. Inset shows late-time behavior on a linear 
scale.
\label{3dHLfig}}
\end{figure}

\subsection{Discussion of Hard-Loop Results}

During the process of instability growth the soft gauge fields get the energy 
for their growth from the hard particles.  In an abelian plasma this energy 
grows exponentially until the energy in the soft field is of the same order of 
magnitude as the energy remaining in the hard particles.  As mentioned above in 
a non-abelian plasma one must rely on numerical simulations due to the presence 
of strong gauge field self-interactions. In Fig.~\ref{3dHLfig} I have plotted 
the time dependence of the energy extracted from the hard particles obtained in 
a 3+1 dimensional simulation of an anisotropic plasma initialized with very weak 
random color noise \cite{Rebhan:2005re}.  As can be seen from this figure at 
$m_\infty t \sim 60$ there is a change from exponential to linear growth with the 
late-time linear slope decreasing as $N_W$ is increased.

The first conclusion that can be drawn from this result is that within non-abelian 
plasmas instabilities will be less efficient at isotropizing the plasma 
than in abelian plasmas.  However, from a theoretical perspective ``saturation'' 
at the soft scale implies that one can still apply the hard-loop effective 
theory self-consistently to understand the behavior of the system at late times. 
Note, however, that the latest simulations 
\cite{Arnold:2005vb,Rebhan:2005re} have only presented results for distributions 
with a finite ${\mathcal O}(1\!\!\rightarrow\!\!10)$ anisotropy and these seem to imply that in 
this case the induced instabilities will not have a significant effect on the hard 
particles.  As a consequence due to the continued expansion of the 
system the anisotropy will increase.  It is therefore important to 
understand the behavior of the system for more extreme anisotropies. 
Additionally, it would be very interesting to study the hard-loop dynamics in an 
expanding system.  Naively, one expects this to change the growth from 
$\exp(\tau)$ to $\exp(\sqrt\tau)$ at short times but there is no clear 
expectation of what will happen in the linear regime.  Significant advances
in this regard have occurred recently \cite{Romatschke:2006wg}.  These
initial results apply to Abelian plasmas and it would be very 
interesting to incorporate expansion in collisionless Boltzmann-Vlasov transport 
in the hard-loop regime and study the late-time behavior in the non-abelian
case.

\section{Wong-Yang-Mills equations}
\label{sec_WYMeqs}

It is also to possible to go beyond the hard-loop approximation and solve
instead the full classical transport equations in three dimensions \cite{Dumitru:2006pz}.
The Vlasov transport
equation for hard gluons with non-Abelian color charge $q^a$ in the
collisionless approximation are~\cite{Wong:1970fu,Heinz:1983nx},
\begin{equation}
 p^{\mu}[\partial_\mu - gq^aF^a_{\mu\nu}\partial^\nu_p
    - gf_{abc}A^b_\mu q^c\partial_{q^a}]f(x,p,q)=0~.   \label{Vlasov}
\end{equation}
Here, $f(t,\bf{x},\bf{p},q^a)$ denotes the single-particle phase space
distribution function.

The Vlasov equation is coupled self-consistently to the Yang-Mills
equation for the soft gluon fields,
\begin{equation}
 D_\mu F^{\mu\nu} = J^\nu = g \int \frac{d^3p}{(2\pi)^3} dq \,q\,
 v^\nu f(t,\bf{x},\bf{p},q)~, \label{YM}
\end{equation}
with $v^\mu\equiv(1,\bf{p}/p)$. These equations reproduce the
``hard thermal loop'' effective action near
equilibrium~\cite{Kelly:1994ig,Kelly:1994dh,Blaizot:1999xk}. However, 
the full classical transport
theory~(\ref{Vlasov},\ref{YM}) also reproduces some higher $n$-point
vertices of the dimensionally reduced effective action for
static gluons~\cite{Laine:2001my} beyond the hard-loop approximation. The
back-reaction of the long-wavelength fields on the hard particles
(``bending'' of their trajectories) is, of course, taken into account,
which is important for understanding particle dynamics in strong
fields.

Eq.~(\ref{Vlasov}) can be solved numerically by replacing
the continuous single-particle distribution $f(\bf{x},\bf{p},q)$
by a large number of test particles:
\begin{equation}
 f(\bf{x},\bf{p},q) = \frac{1}{N_{\rm test}}\sum_i 
  \delta(\bf{x}-\bf{x}_i(t)) \,(2\pi)^3 \delta(\bf{p}-\bf{p}_i(t)) \,
   \delta(q^a-q_i^a(t))~,  \label{TestPartAnsatz}
\end{equation}
where $\bf{x}_i(t)$, $\bf{p}_i(t)$ and ${q}_i^a(t)$ are the
coordinates of an individual test particle and $N_{\rm test}$ denotes
the number of test-particles per physical particle. The {\sl
Ansatz}~(\ref{TestPartAnsatz}) leads to Wong's
equations~\cite{Wong:1970fu,Heinz:1983nx}
\begin{eqnarray}
\frac{d\bf{x}_i}{dt} &=& \bf{v}_i,\\ \frac{d\bf{p}_i}{dt} &=& g\,
q_i^a \,
\left( \bf{E}^a + \bf{v}_i \times \bf{B}^a \right),\label{pdot}\\
\frac{d\bf{q}_i}{dt} &=& ig\, v^{\mu}_i \, [ A_\mu, \bf{q}_i],\\ 
J^{a\,\nu} &=& \frac{g}{N_{\rm test}} \sum_i q_i^a \,v^\nu
\,\delta(\bf{x}-\bf{x}_i(t)).
\end{eqnarray}
for the $i$-th test particle.  The time evolution of
the Yang-Mills field can be followed by the standard Hamiltonian
method~\cite{Ambjorn:1990pu} in $A^0=0$ gauge.  For details of the 
numerical implementation used see Ref.~\cite{Dumitru:2006pz}.

\begin{figure}[t]
\vspace{4mm}
\includegraphics[width=7cm]{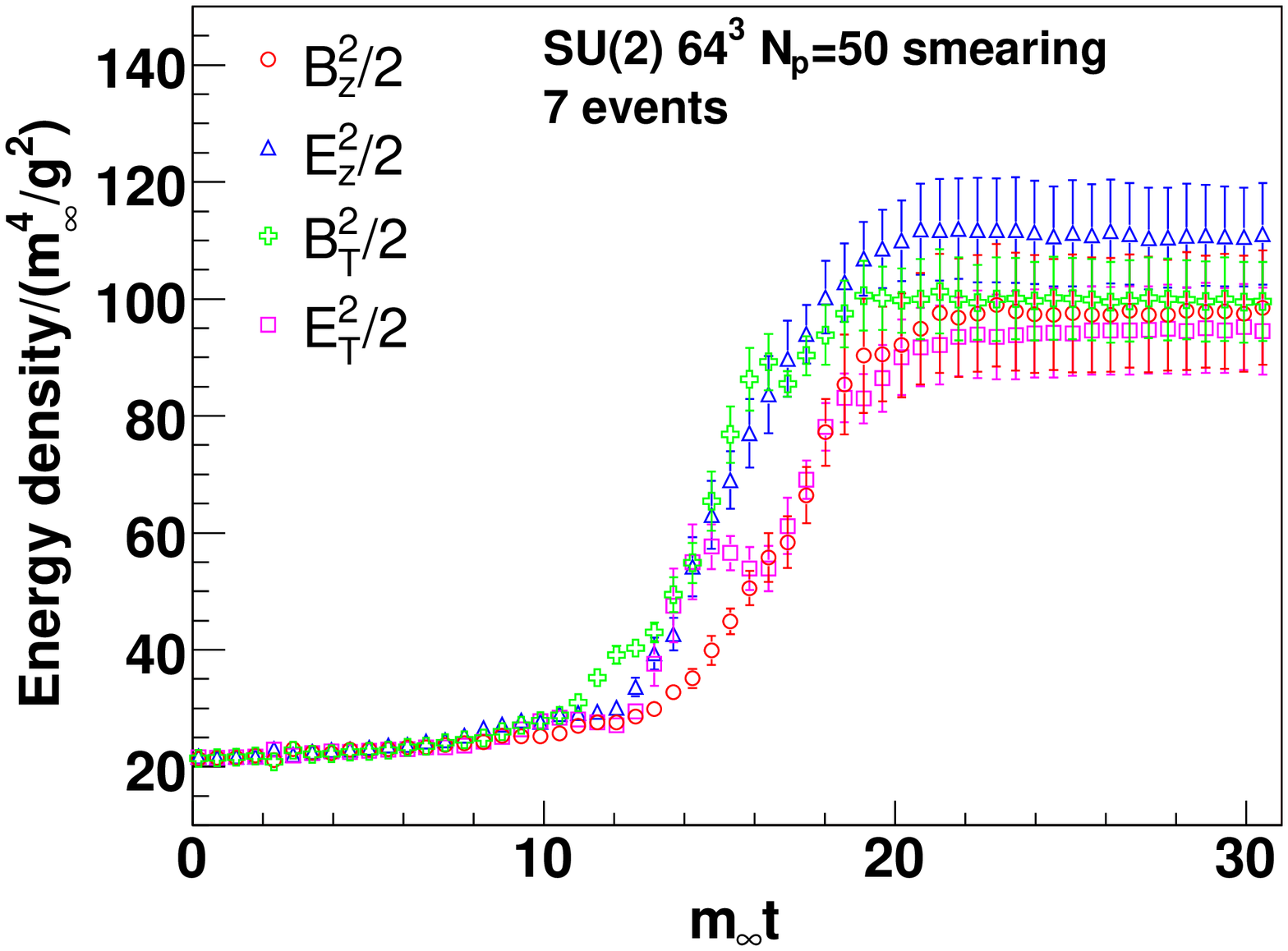}
\includegraphics[width=10cm]{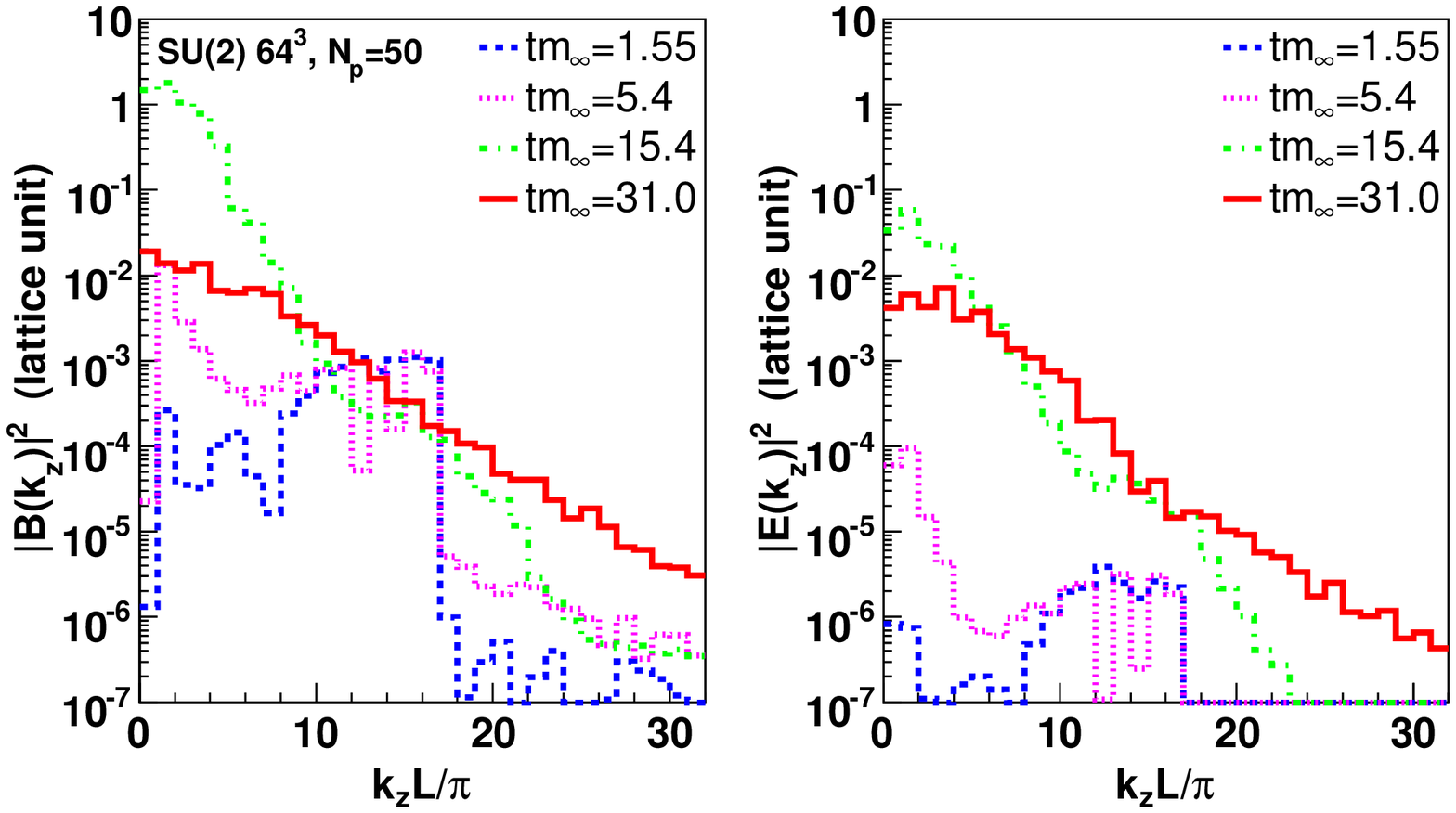}
\vspace{-7mm}
\caption{
Leftmost figure shows the time evolution of the field energy densities for $SU(2)$
gauge group resulting from a highly anisotropic initial particle
momentum distribution.  Simulation parameters are $L=5$~fm, $p_{\rm
hard}=16$~GeV, $g^2\,n_g=10/$fm$^3$, $m_\infty=0.1$~GeV.
Right two panels show the corresponding fourier transforms of the
electric and magnetic fields at different times which are indicated in
the legend.}
\vspace{-4mm}
\label{cpicfig}
\end{figure}

In Fig.~\ref{cpicfig} I present the results of the three-dimensional
simulations published in Ref.~\cite{Dumitru:2006pz}.  The leftmost panel shows
the time evolution of the field energy densities for $SU(2)$
gauge group resulting from a highly anisotropic initial particle
momentum distribution.  The right two panels show the corresponding fourier 
transforms of the electric and magnetic fields at different times which are 
indicated in the legend.  

The behavior shown in Fig.~\ref{cpicfig} indicates that the results obtained
from the hard-loop simulations and direct numerical solution of the Wong-Yang-Mills
(WYM) equations are qualitatively similar in that both show that for non-abelian
gauge theories there is a saturation of the energy transferred to the soft modes by
the gauge instability.  Additionally, as can be seen from the fourier transforms
in right two panels of Fig.~\ref{cpicfig} the saturation is accompanied by an
``avalanche'' of energy transferred to soft field modes to higher frequency
field modes with saturation occurring when the hardest lattice modes are filled.
A more thorough analytic understanding of this ultraviolet avalanche is lacking
at this point in time although some advances in this regard have been made
recently \cite{Mueller:2006up}.  Additionally, since within the numerical solution
of the WYM equations the ultraviolet modes become populated rapidly this means
that the effective theory which relies on a separation between hard (particle)
and soft (field) scales breaks down.  This should motivate research into numerical 
methods which can be used to ``shuffle'' field modes to particles when their 
momentum becomes too large and vice-versa for hard particles.  This is a difficult
task but work on such algorithms is in progress.  Hopefully, using such methods it will
be possible to simulate the non-equilibrium dynamics of anisotropic plasmas in
a self-consistent numerical framework which treats particles and fields and
the transmutation among these two types of degrees of freedom smoothly.

\section{Outlook}

An obviously important question is whether quark-gluon plasma instabilities 
and/or the physics of anisotropic plasmas in general play an important 
phenomenological role at RHIC or LHC energies.  In this regard the recent papers 
of 
Refs.~\cite{Romatschke:2003vc,Romatschke:2004au,Schenke:2006fz,Romatschke:2006bb} 
provide theoretical frameworks which can be used to calculate the impact of 
anisotropic momentum-space distributions on observables
such as jet shapes and the rapidity dependence of medium-produced photons.


\bibliographystyle{h-elsevier}
\bibliography{strickland}


\end{document}